\documentclass[%
 reprint,
 superscriptaddress,
 preprintnumbers,
 amsmath,amssymb,
 aps,
 floatfix,
]{revtex4-2}

\usepackage{graphicx}
\usepackage{dcolumn}
\usepackage{bm}
\usepackage{xcolor}

\usepackage{lineno}

\usepackage{siunitx}
\newcommand{\ethr}{\left<E_\mathrm{thr}\cos\theta\right>}

\begin{document}

\title{Seasonal Variation of Multiple-Muon Cosmic Ray Air Showers Observed in the NOvA Detector on the Surface}

\preprint{FERMILAB-PUB-21-224-ND}

\newcommand{\ANL}{Argonne National Laboratory, Argonne, Illinois 60439, 
USA}
\newcommand{\ICS}{Institute of Computer Science, The Czech 
Academy of Sciences, 
182 07 Prague, Czech Republic}
\newcommand{\IOP}{Institute of Physics, The Czech 
Academy of Sciences, 
182 21 Prague, Czech Republic}
\newcommand{\Atlantico}{Universidad del Atlantico,
Carrera 30 No. 8-49, Puerto Colombia, Atlantico, Colombia}
\newcommand{\BHU}{Department of Physics, Institute of Science, Banaras 
Hindu University, Varanasi, 221 005, India}
\newcommand{\UCLA}{Physics and Astronomy Department, UCLA, Box 951547, Los 
Angeles, California 90095-1547, USA}
\newcommand{\Caltech}{California Institute of 
Technology, Pasadena, California 91125, USA}
\newcommand{\Cochin}{Department of Physics, Cochin University
of Science and Technology, Kochi 682 022, India}
\newcommand{\Charles}
{Charles University, Faculty of Mathematics and Physics,
 Institute of Particle and Nuclear Physics, Prague, Czech Republic}
\newcommand{\Cincinnati}{Department of Physics, University of Cincinnati, 
Cincinnati, Ohio 45221, USA}
\newcommand{\CSU}{Department of Physics, Colorado 
State University, Fort Collins, CO 80523-1875, USA}
\newcommand{\CTU}{Czech Technical University in Prague,
Brehova 7, 115 19 Prague 1, Czech Republic}
\newcommand{\Dallas}{Physics Department, University of Texas at Dallas,
800 W. Campbell Rd. Richardson, Texas 75083-0688, USA}
\newcommand{\DallasU}{University of Dallas, 1845 E 
Northgate Drive, Irving, Texas 75062 USA}
\newcommand{\Delhi}{Department of Physics and Astrophysics, University of 
Delhi, Delhi 110007, India}
\newcommand{\JINR}{Joint Institute for Nuclear Research,  
Dubna, Moscow region 141980, Russia}
\newcommand{\Erciyes}{
Department of Physics, Erciyes University, Kayseri 38030, Turkey}
\newcommand{\FNAL}{Fermi National Accelerator Laboratory, Batavia, 
Illinois 60510, USA}
\newcommand{\UFG}{Instituto de F\'{i}sica, Universidade Federal de 
Goi\'{a}s, Goi\^{a}nia, Goi\'{a}s, 74690-900, Brazil}
\newcommand{\Guwahati}{Department of Physics, IIT Guwahati, Guwahati, 781 
039, India}
\newcommand{\Harvard}{Department of Physics, Harvard University, 
Cambridge, Massachusetts 02138, USA}
\newcommand{\Houston}{Department of Physics, 
University of Houston, Houston, Texas 77204, USA}
\newcommand{\IHyderabad}{Department of Physics, IIT Hyderabad, Hyderabad, 
502 205, India}
\newcommand{\Hyderabad}{School of Physics, University of Hyderabad, 
Hyderabad, 500 046, India}
\newcommand{\IIT}{Illinois Institute of Technology,
Chicago IL 60616, USA}
\newcommand{\Indiana}{Indiana University, Bloomington, Indiana 47405, 
USA}
\newcommand{\INR}{Institute for Nuclear Research of Russia, Academy of 
Sciences 7a, 60th October Anniversary prospect, Moscow 117312, Russia}
\newcommand{\Iowa}{Department of Physics and Astronomy, Iowa State 
University, Ames, Iowa 50011, USA}
\newcommand{\Irvine}{Department of Physics and Astronomy, 
University of California at Irvine, Irvine, California 92697, USA}
\newcommand{\Jammu}{Department of Physics and Electronics, University of 
Jammu, Jammu Tawi, 180 006, Jammu and Kashmir, India}
\newcommand{\Lebedev}{Nuclear Physics and Astrophysics Division, Lebedev 
Physical 
Institute, Leninsky Prospect 53, 119991 Moscow, Russia}
\newcommand{\Magdalena}{Universidad del Magdalena, Carrera 32 No 22-08 Santa Marta, Colombia}
\newcommand{\MSU}{Department of Physics and Astronomy, Michigan State 
University, East Lansing, Michigan 48824, USA}
\newcommand{\Crookston}{Math, Science and Technology Department, University 
of Minnesota Crookston, Crookston, Minnesota 56716, USA}
\newcommand{\Duluth}{Department of Physics and Astronomy, 
University of Minnesota Duluth, Duluth, Minnesota 55812, USA}
\newcommand{\Minnesota}{School of Physics and Astronomy, University of 
Minnesota Twin Cities, Minneapolis, Minnesota 55455, USA}
\newcommand{\Mississippi}{University of Mississippi, University, Mississippi 38677, USA}
\newcommand{\NISER}{National Institute of Science Education and Research,
Khurda, 752050, Odisha, India}
\newcommand{\Oxford}{Subdepartment of Particle Physics, 
University of Oxford, Oxford OX1 3RH, United Kingdom}
\newcommand{\Panjab}{Department of Physics, Panjab University, 
Chandigarh, 160 014, India}
\newcommand{\Pitt}{Department of Physics, 
University of Pittsburgh, Pittsburgh, Pennsylvania 15260, USA}
\newcommand{\QMU}{School of Physics and Astronomy,
 Queen Mary University of London,
London E1 4NS, United Kingdom}
\newcommand{\RAL}{Rutherford Appleton Laboratory, Science 
and 
Technology Facilities Council, Didcot, OX11 0QX, United Kingdom}
\newcommand{\SAlabama}{Department of Physics, University of 
South Alabama, Mobile, Alabama 36688, USA} 
\newcommand{\Carolina}{Department of Physics and Astronomy, University of 
South Carolina, Columbia, South Carolina 29208, USA}
\newcommand{\SDakota}{South Dakota School of Mines and Technology, Rapid 
City, South Dakota 57701, USA}
\newcommand{\SMU}{Department of Physics, Southern Methodist University, 
Dallas, Texas 75275, USA}
\newcommand{\Stanford}{Department of Physics, Stanford University, 
Stanford, California 94305, USA}
\newcommand{\Sussex}{Department of Physics and Astronomy, University of 
Sussex, Falmer, Brighton BN1 9QH, United Kingdom}
\newcommand{\Syracuse}{Department of Physics, Syracuse University,
Syracuse NY 13210, USA}
\newcommand{\Tennessee}{Department of Physics and Astronomy, 
University of Tennessee, Knoxville, Tennessee 37996, USA}
\newcommand{\Texas}{Department of Physics, University of Texas at Austin, 
Austin, Texas 78712, USA}
\newcommand{\Tufts}{Department of Physics and Astronomy, Tufts University, Medford, 
Massachusetts 02155, USA}
\newcommand{\UCL}{Physics and Astronomy Department, University College 
London, 
Gower Street, London WC1E 6BT, United Kingdom}
\newcommand{\Virginia}{Department of Physics, University of Virginia, 
Charlottesville, Virginia 22904, USA}
\newcommand{\WSU}{Department of Mathematics, Statistics, and Physics,
 Wichita State University, 
Wichita, Kansas 67206, USA}
\newcommand{\WandM}{Department of Physics, William \& Mary, 
Williamsburg, Virginia 23187, USA}
\newcommand{\Wisconsin}{Department of Physics, University of 
Wisconsin-Madison, Madison, Wisconsin 53706, USA}
\newcommand{\deceased}{Deceased.}
\affiliation{\ANL}
\affiliation{\Atlantico}
\affiliation{\BHU}
\affiliation{\Caltech}
\affiliation{\Charles}
\affiliation{\Cincinnati}
\affiliation{\Cochin}
\affiliation{\CSU}
\affiliation{\CTU}
\affiliation{\Delhi}
\affiliation{\Erciyes}
\affiliation{\FNAL}
\affiliation{\UFG}
\affiliation{\Guwahati}
\affiliation{\Harvard}
\affiliation{\Houston}
\affiliation{\Hyderabad}
\affiliation{\IHyderabad}
\affiliation{\IIT}
\affiliation{\Indiana}
\affiliation{\ICS}
\affiliation{\INR}
\affiliation{\IOP}
\affiliation{\Iowa}
\affiliation{\Irvine}
\affiliation{\JINR}
\affiliation{\Lebedev}
\affiliation{\Magdalena}
\affiliation{\MSU}
\affiliation{\Duluth}
\affiliation{\Minnesota}
\affiliation{\Mississippi}
\affiliation{\NISER}
\affiliation{\Panjab}
\affiliation{\Pitt}
\affiliation{\QMU}
\affiliation{\SAlabama}
\affiliation{\Carolina}
\affiliation{\SMU}
\affiliation{\Stanford}
\affiliation{\Sussex}
\affiliation{\Syracuse}
\affiliation{\Texas}
\affiliation{\Tufts}
\affiliation{\UCL}
\affiliation{\Virginia}
\affiliation{\WSU}
\affiliation{\WandM}
\affiliation{\Wisconsin}

\author{M.~A.~Acero}
\affiliation{\Atlantico}

\author{P.~Adamson}
\affiliation{\FNAL}



\author{L.~Aliaga}
\affiliation{\FNAL}






\author{N.~Anfimov}
\affiliation{\JINR}


\author{A.~Antoshkin}
\affiliation{\JINR}


\author{E.~Arrieta-Diaz}
\affiliation{\Magdalena}

\author{L.~Asquith}
\affiliation{\Sussex}


\author{A.~Aurisano}
\affiliation{\Cincinnati}


\author{A.~Back}
\affiliation{\Iowa}

\author{C.~Backhouse}
\affiliation{\UCL}

% ...

\collaboration{The NOvA Collaboration}
\noaffiliation

\date{\today}

\begin{abstract}

We report the rate of cosmic ray air showers
with multiplicities exceeding 15 muon tracks recorded in the NOvA Far Detector between May 2016 and May 2018.
The detector is located on the surface under an overburden 
of \SI{3.6}{meters} water equivalent.
We observe a seasonal dependence in the rate of multiple-muon showers,
which varies in magnitude with multiplicity and zenith angle.
During this period, the effective atmospheric temperature and surface pressure
ranged between \SIrange{210}{230}{\kelvin} and \SIrange{940}{990}{mbar}, respectively;
the shower rates are anti-correlated with the variation in the effective 
temperature.
The variations are about
30\% larger for the highest multiplicities 
than the lowest multiplicities and 20\% 
larger for showers near the horizon than vertical showers.

\end{abstract}

\maketitle

\section{Introduction}
\label{sec:intro}

\begin{figure*}[t]
    \centering
    \includegraphics[width=0.95\textwidth]{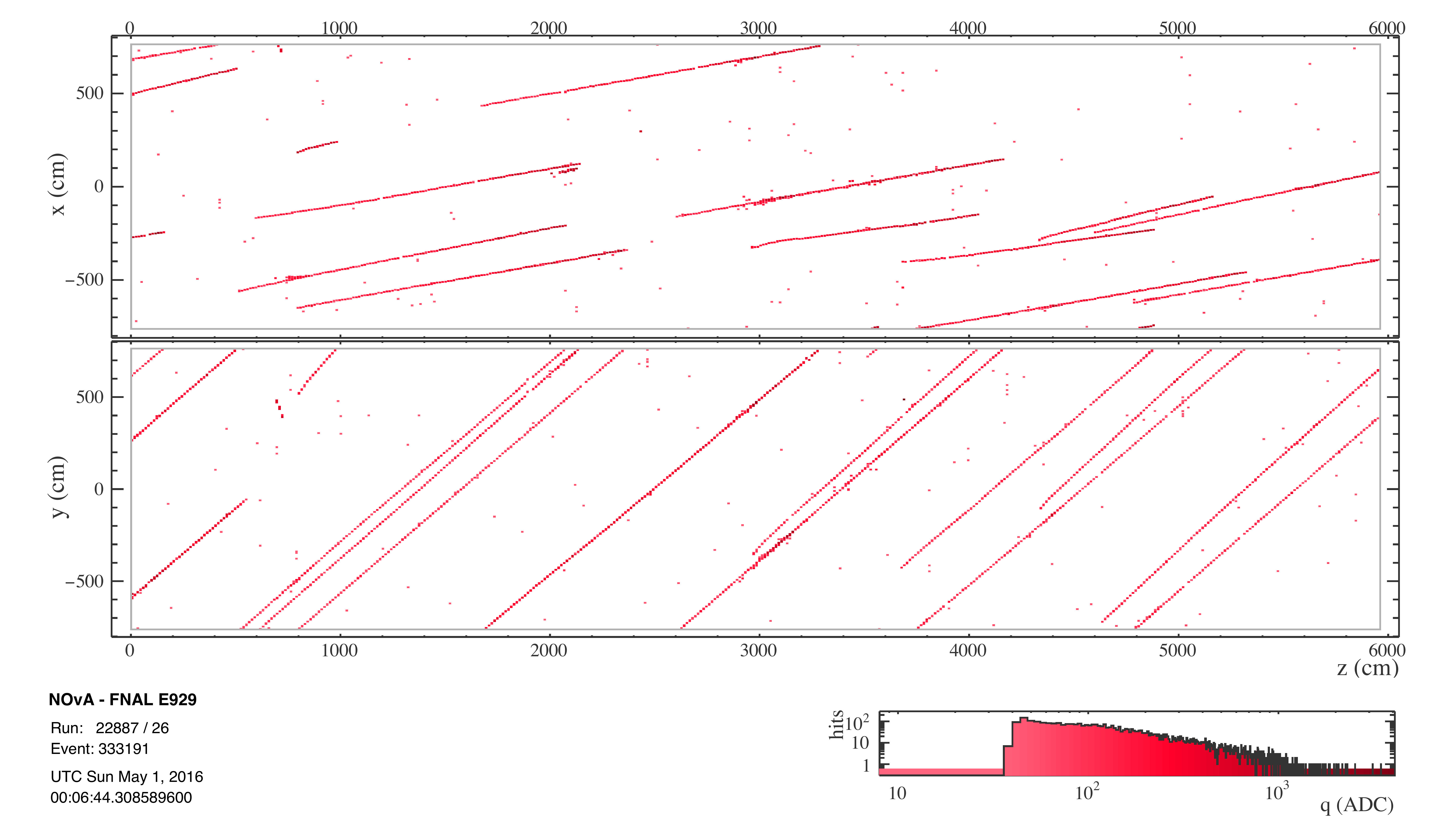}
    \caption{A multiple-muon air shower recorded in the NOvA detector.
    The top is a view of the event from above the detector and the bottom is a view from the side.
    Each linear set of hits is one muon traversing the detector.
    The color corresponds to how much energy was deposited in each detector cell in units of ADC counts shown in the bottom right histogram.}
    \label{fig:evd}
\end{figure*}

Several experiments have observed seasonality in the rate of muons 
from cosmic ray air showers which depends on the density 
profile of the earth's atmosphere.
As the density changes, the relative numbers 
of pions and kaons which interact or decay in the
air showers change, altering the observed rate of muons.
During the summer months, when temperature is high and the density is lowest,
meson decays are more probable, which leads to an 
expected peak in the muon rates.
This summer peak has been confirmed by many experiments in 
single-muon air 
showers~\cite{dayabay,carpet,minossingle,opera} 
and is explained by existing models~\cite{teff}.

However, existing models fail to fully explain
the observations of multiple-muon air showers.
The DECOR~\cite{decor} and GRAPES~\cite{grapes} experiments have 
reported peak rates of multiple-muon showers in the winter in 
detectors close to the surface, opposite to expectations 
and observations for single-muons.
DECOR attributed its observation to geometric effects arising from altitude 
differences in meson production.
The MINOS experiment~\cite{minos} observed a winter peak in two underground detectors, with minimum muon energies of \SI{60}{GeV} and \SI{700}{GeV}, 
and showed that at a depth of at 
least \SI{225}{meters} water equivalent the 
effect from altitude differences suggested by 
DECOR was too small to fully explain the observation.
NOvA also previously reported a peak rate in winter of 
multiple-muon cosmic ray air showers using its Near Detector~\cite{ndpaper} 
located at the same depth as one of the MINOS detectors with a threshold energy of \SI{60}{GeV}.

The NOvA Far Detector is located near the surface where no seasonal variation is expected for low energy, single-muon air showers~\cite{stefano}.
However, the Far Detector has a top surface area 
which is 15 times larger than the Near Detector 
making it sensitive to much higher multiplicity showers.

In this paper, we report the observation of a winter maximum
of multiple-muon air showers using NOvA's Far Detector.
Since no quantitative models for multiple-muon air showers reproduce the effects we observe, the seasonal effect will be quantified using two different methods.
First, the rate of multiple-muon air showers is compared to the temperature and surface pressure of the atmosphere above the detector site and, second, by fitting the observed muon rate to a cosine function.
We also show how the strength of this observation varies
with observed muon multiplicity and arrival direction 
in a surface detector for the first time.

\section{The NOvA Far Detector}
\label{sec:det}

The NOvA Far Detector is a 
\SI{14}{kt} sampling calorimeter, \SI{15.5x15.5x59.8}{\meter} in size, segmented into \SI{4x6x15.5}{cm} channels.
The channels are arranged into alternating horizontal and vertical planes.
The detector was designed to detect neutrino 
interactions in the NuMI beam from Fermi National 
Accelerator Laboratory~\cite{ana2019}.
It is located on the surface in Minnesota 
near the U.S.-Canada Border at 
$\left(48.4^\circ~\mathrm{N},92.8^\circ~\mathrm{W}\right)$.
The detector has been operating with more than 97\% up-time efficiency since 2014. 
This analysis samples 15\% of the total cosmic ray data set collected between 
May 2016 and May 2018. The detector design and detection 
mechanism are described in~\cite{NOVATDR}.

The detector sits just below the surface level beneath an overburden to shield the detector from cosmic ray photons and electrons.
It consists of \SI{1.2}{\meter} of 
concrete and \SI{15}{cm} of barite rock 
giving a total of \SI{3.6}{meters} water equivalent.
Three sides of the building are surrounded by a sloped berm of granite rock at $30^\circ$ to the surface.
This shielding is not present north of the detector where the detector assembly hall is located.
Above the horizon, this overburden adds an additional muon energy threshold of 
$\ethr\approx\SI{1.5}{GeV}$, where $\theta$ is the zenith angle, to reach the detector.
On average, 10 billion cosmic ray muons traverse the detector each day.

Data from multiple-muon showers are recorded for analysis~\cite{novasn} 
whenever the detector records total visible energy 
in excess 
of approximately \SI{20}{GeV} of energy deposited
in a \SI{50}{\micro\second} readout
distributed among at least 120 
of the detector's total of \SI{343968}{channels}.
A typical muon with $\theta=\SI{30}{\degree}$ traversing the center of the detector will deposit around \SI{2.5}{GeV} of visible energy.

\section{Atmospheric and Muon Data}

The signal of muon air shower events in the 
detector is a large number of coincident, parallel tracks.
Fig.~\ref{fig:evd} shows the signal topology of a multiple-muon shower recorded in the detector.
Reconstruction of these showers
begins by isolating the time range containing the activity
of interest from other detector activity and 
suppressing isolated detector hits which do not contribute to tracks.
A Hough transform determines the overall shower 
angle in each view of the detector.
These angles seed the construction of individual muon tracks.
These algorithms produce a 
zenith angle, azimuthal angle, and 
multiplicity assignment for each air shower.
The multiplicity reported here is the observed multiplicity within the detector.
Because air showers can be much larger than the surface area of the NOvA detector, no attempt was made to estimate the true multiplicity.

The reconstruction was optimized and validated using 
air-shower simulations based on 
COsmic Ray SImulations for KAscade (CORSIKA)~\cite{corsika} 
with a range of primary cosmic ray energies in order to explore 
performance on showers with a variety of multiplicities.
The reconstructed multiplicity is within $\pm1$ of the true 
multiplicity in the detector for 70\% of showers and within $\pm4$ for 95\% of showers.
The most common reconstruction failure is due to muon tracks that are nearly overlapping in space in one view and are treated as a single track.

\begin{figure}[t]
    \centering
    \includegraphics[width=0.48\textwidth]{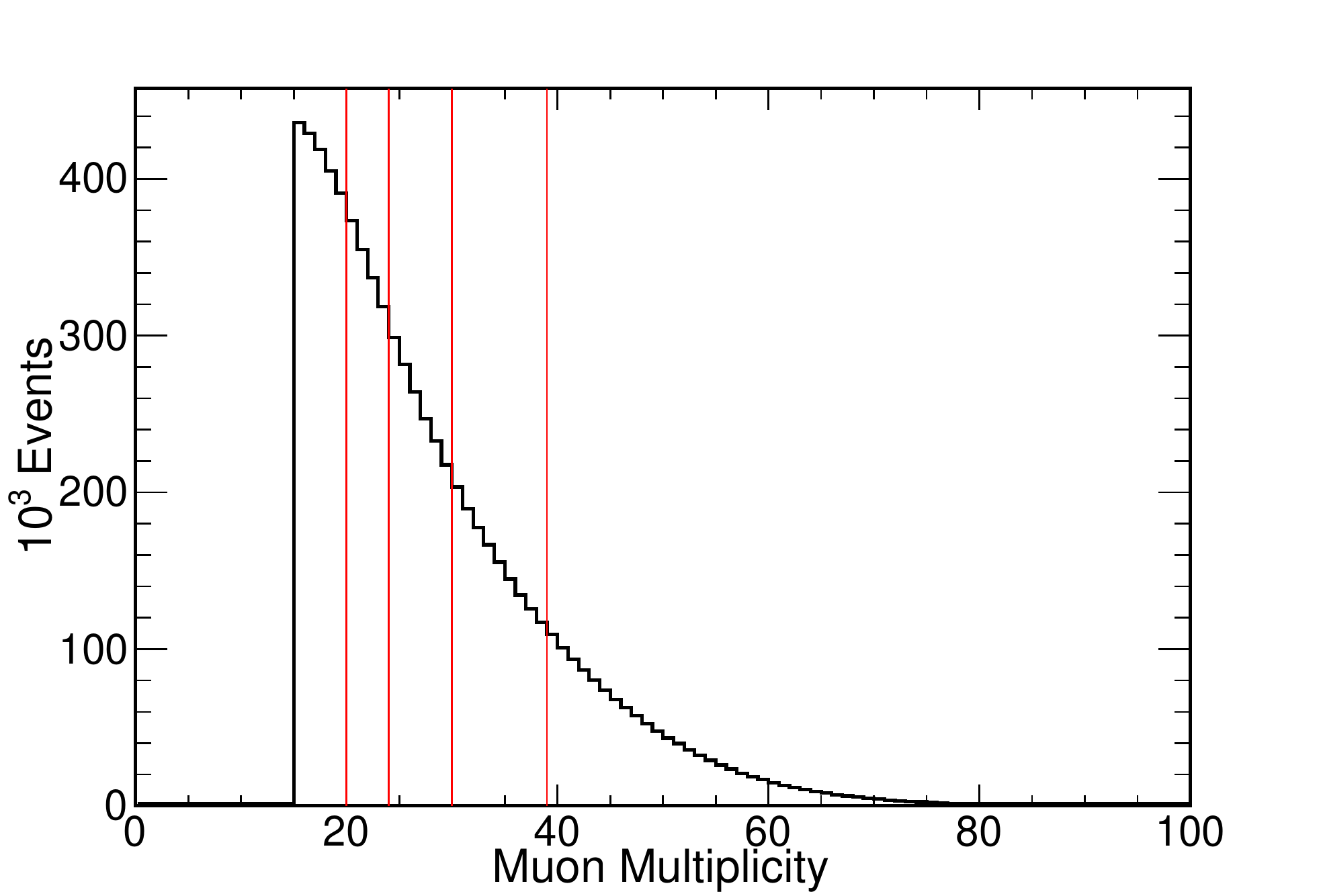}
    \includegraphics[width=0.48\textwidth]{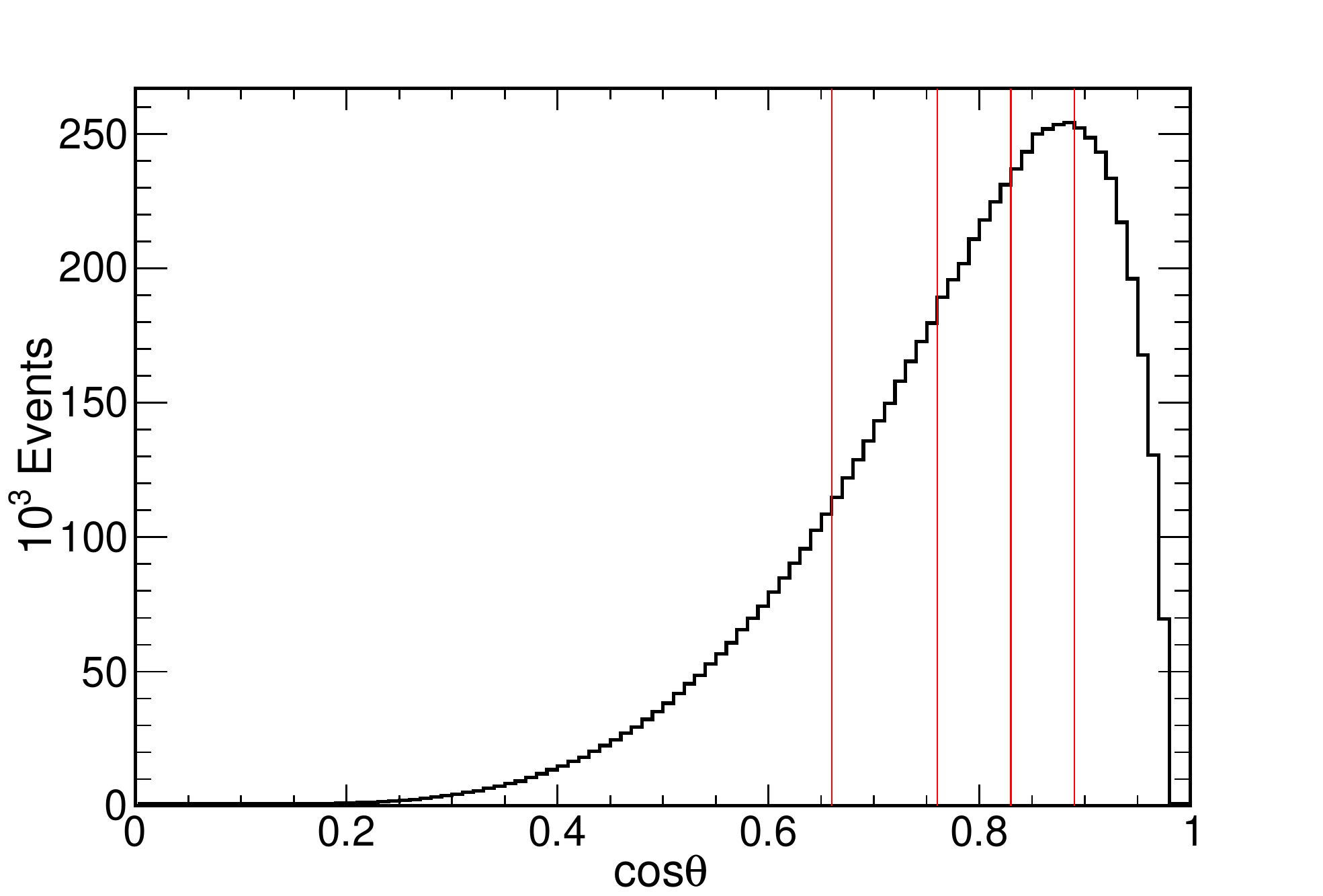}
    \caption{Reconstructed quantities for all showers used in the analysis.
    Top: The muon multiplicity of air showers recorded in the NOvA detector.
    Only multiplicities of 15 or more are considered.
    Bottom: The cosine of the zenith angles of air showers.
    Showers with vertical zenith angle or horizontal showers from the east/west are removed from the analysis due to reconstruction challenges which suppresses the showers at high $\cos\theta$.
    In both panels, the vertical lines denote the bin boundaries used in section~\ref{sec:obs}.}
    \label{fig:showers}
\end{figure}

We apply a number of data selection 
requirements to ensure uniform detector acceptance for air showers.
While it is possible for the NOvA detectors to operate with only a subset of 
components active, we only analyze data-taking 
periods when the detector was completely active
to ensure continuity of the muon tracks.
Air showers travelling nearly parallel to the 
detector planes (up-down, east-west) are removed as the reconstruction
cannot produce complete tracks for these orientations.
Candidate events with very large energy 
deposits per hit likely contain large hadron showers from the overburden.
These events are either not associated with air 
showers or hinder our ability to reconstruct the 
air shower direction or multiplicity and are 
removed from the sample.
Showers with a 
reconstructed multiplicity less than 15 are removed to 
avoid trigger inefficiencies at
low multiplicities and to ensure a uniform 
efficiency over the analysis sample.
From a CORSIKA simulation, the typical primary energy to make 15 muons in the detector is \SIrange{30}{100}{TeV}.
The reconstructed multiplicity and zenith angle of all selected
showers can be seen in Fig.~\ref{fig:showers}.

\begin{figure}[t]
    \centering
    \includegraphics[width=0.48\textwidth]{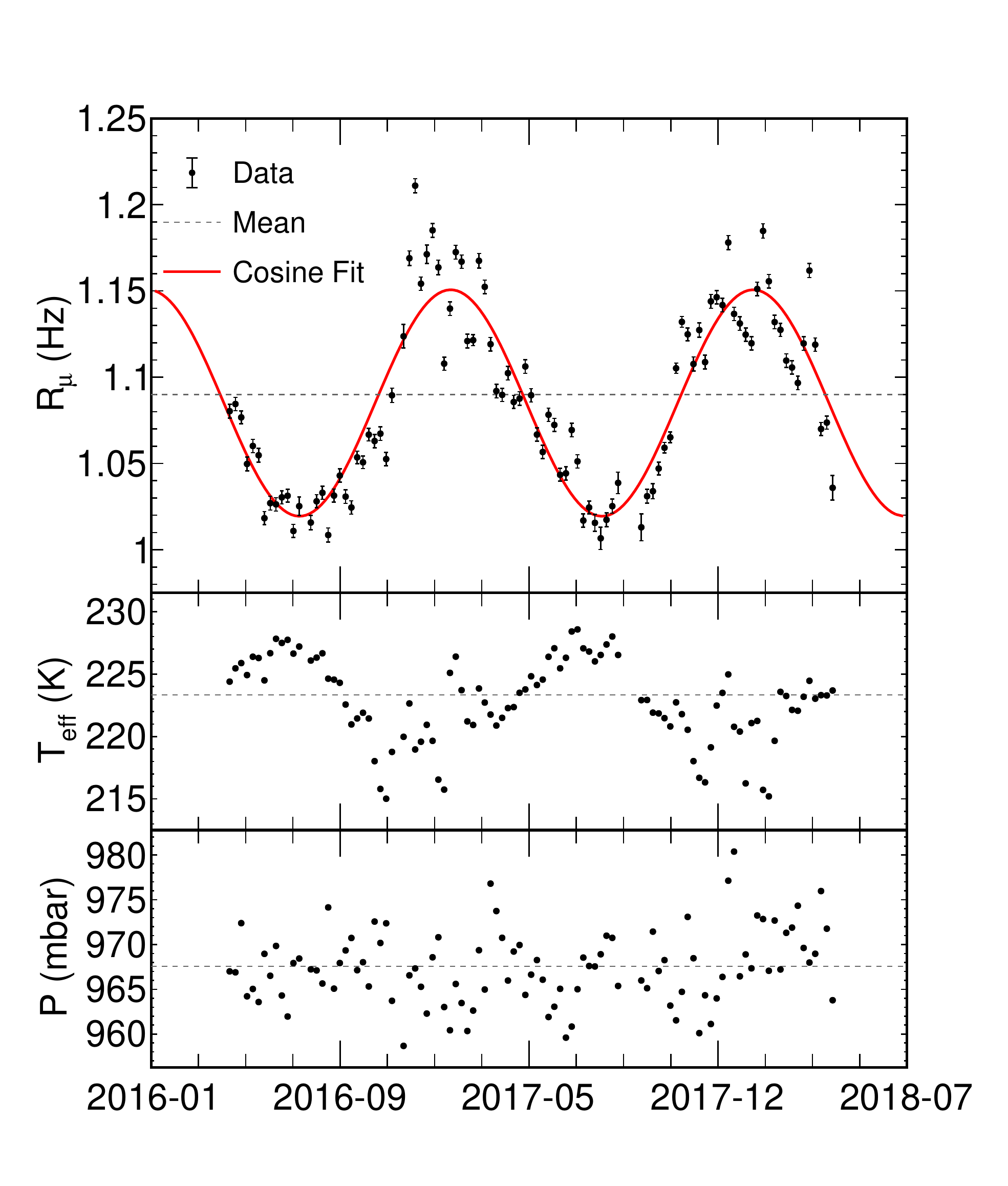}
    \caption{Data recorded at the detector site over time.
    Each point represents the average over a week of data.
    Top: The rate of multiple-muon air showers with an average rate of $\left<R_\mu\right>=\SI{1.09}{Hz}$.
    The red curve shows a sinusoidal fit to the data.
    Middle: The effective temperature with an average of $\left<T_\mathrm{eff}\right>=\SI{223}{\kelvin}$.
    The errors are too small to display.
    Bottom: The surface pressure with an average value of $\langle P\rangle=\SI{968}{mbar}$.
    The errors are too small to display.
    No clear seasonal trend is observed in the pressure.}
    \label{fig:seasonal}
\end{figure}

The livetime used to compute the shower rates 
is recorded by data acquisition 
processes, which monitor the trigger data streams.
Fig.~\ref{fig:seasonal} shows the rate of 
cosmic ray air showers for the full analysis period.
The rates reach their maximum values during the winter months.
In total, $7.64\times10^6$ multiple-muon 
showers with an average multiplicity of 28 are analyzed.
The showers have an average rate of $\left<R_\mu\right>=\SI{1.09}{Hz}$.


The measured rate of multiple-muon showers will be compared to atmospheric conditions.
We use atmospheric data provided by the European Centre for 
Medium-Range Weather Forecasts (ECMWF)~\cite{ecmwf}.
These global data are provided four times per day with a 
resolution of \SI{0.75}{\degree} in latitude and longitude.
In this analysis, we average the temperature at the four points nearest to the 
detector location that are closest in time to the air shower.

The effective temperature of the atmosphere above the detector
is a weighted average of the temperature recorded at pressure levels 
ranging from \SIrange{1}{1000}{hPa} 
over the depth of the atmosphere with higher weight given to altitudes with muon production~\cite{teff}.
This model is only valid for muons from leading pions and kaons produced in the primary interaction and is applicable for single-muon events.
However, this computation approximates what the effective temperature would be for multiple-muon events.
The average effective temperature for this analysis 
is $\left<T_\mathrm{eff}\right>=\SI{223}{\kelvin}$.
The surface pressure data is also reported with an 
average value of $\langle P\rangle=\SI{968}{mbar}$.
Fig.~\ref{fig:seasonal} shows the variations in these quantities 
over the analysis period.

\section{Seasonal Effects}

Fig.~\ref{fig:seasonal} shows a clear seasonal variation in the 
rate of multiple-muon air showers with peak rates in both winters.
We employ two methods to quantify the significance of these peaks.

The first method correlates the atmospheric 
temperature and the muon rates.
We compute a temperature correlation coefficient, $\alpha_\mathrm{T}$~\cite{teff}:
\begin{linenomath*}
\begin{equation}
    \frac{\Delta R_\mu}{\langle R_\mu\rangle}=\alpha_\mathrm{T}\frac{\Delta T_\mathrm{eff}}{\langle T_\mathrm{eff}\rangle},
\end{equation}
\end{linenomath*}
where $\Delta R_\mu$ is the difference from the mean rate of multiple-muon showers and similarly for the effective temperature, $T_\mathrm{eff}$.
This effective temperature model 
has been shown to be closely correlated
with the rate of single-muon showers.
The value of $\alpha_\mathrm{T}$ is dependent 
on the threshold energy of 
detected muons and thus on the depth of the detector.
Surface detectors are expected to exhibit no 
temperature dependence with $\alpha_\mathrm{T}\approx0$ for 
single-muon showers.
However, the model does not 
accurately explain the development of 
multiple-muon showers~\cite{stefano} 
where many competing effects 
contribute to the observed rate.
Despite this limitation, the magnitude of $\alpha_\mathrm{T}$ 
still demonstrates a correlation 
between temperature and observed multiple-muon rate.
Applying a linear fit to the multiple-muon rates as a function of the 
effective temperatures from Fig.~\ref{fig:seasonal}, results in a temperature correlation coefficient 
of $\alpha_\mathrm{T}=-1.14\pm0.02$.

The atmospheric pressure at the surface can affect the survival probability of low-energy muons as 
they approach the detector and alter the observed rate~\cite{muondecay}.
The barometric coefficient, $\beta$, is measured by:
\begin{linenomath*}
\begin{equation}
    \frac{\Delta R_\mu}{\langle R_\mu\rangle}=\beta\Delta P,
\end{equation}
\end{linenomath*}
where $\Delta P$ is difference in 
pressure from the mean.
The barometric coefficient has been 
shown to accurately relate rates of 
single-muon showers to the pressure~\cite{carpet}.
A fit between the multiple-muon rates and the surface pressures from 
Fig.~\ref{fig:seasonal}, yields a barometric correlation coefficient of $\beta=\SI[separate-uncertainty = true]{-0.08(1)}{\%/mbar}$.

\begin{figure}[t]
    \centering
    \includegraphics[width=0.48\textwidth]{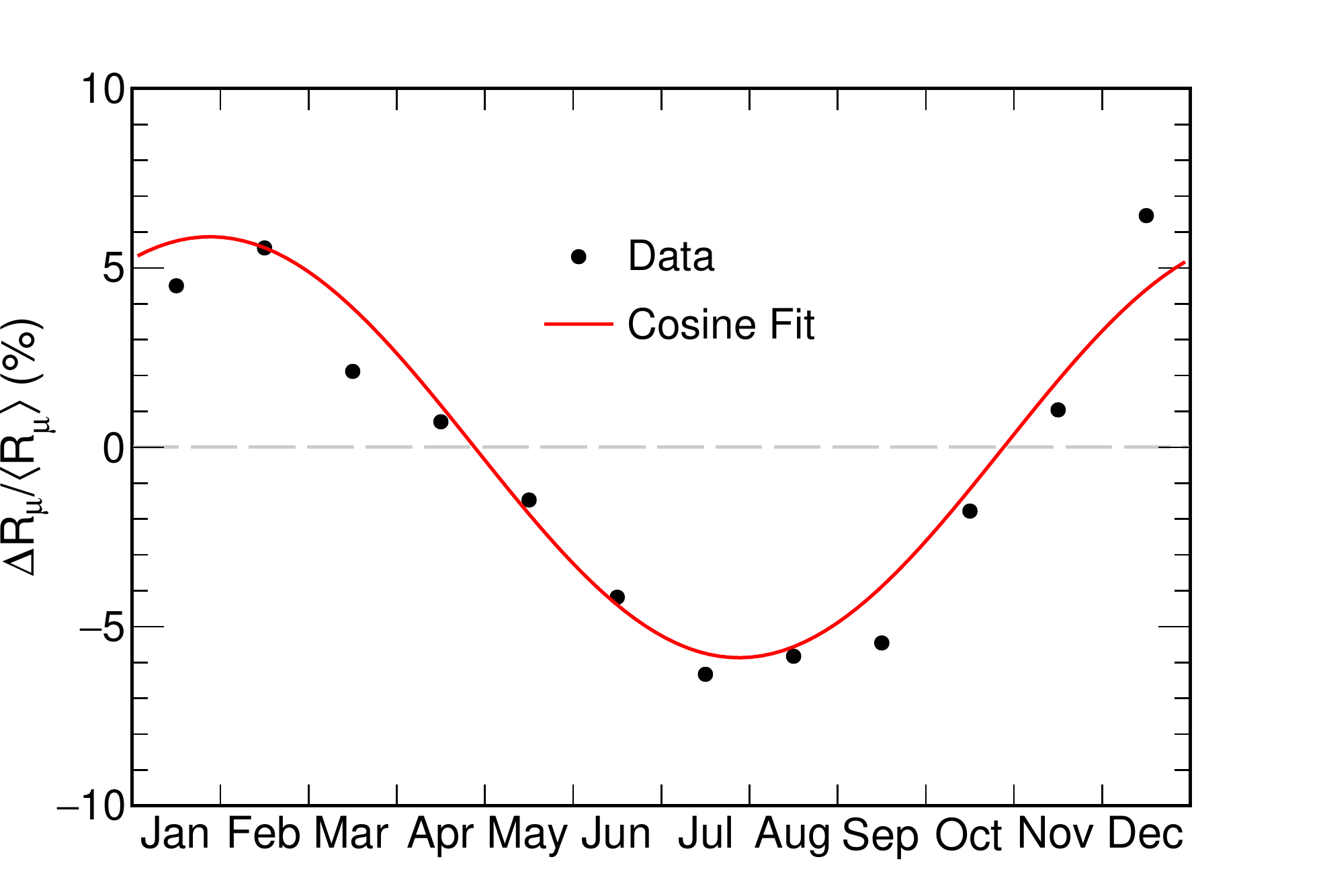}
    \caption{The percent change in the multiple-muon rate.
    The two years of data is averaged to give the rate as a function of month of year. The errors are too small to display.}
    \label{fig:moy}
\end{figure}

The value of $\alpha_\mathrm{T}$ is 
statistically different from zero and negative, 
as expected for an anti-correlation 
between temperature and multiple-muon shower rates.
$\beta$ is also negative, which is as 
expected for the single-muon case for a 
surface detector~\cite{carpet}.
However, the relationship between the 
pressure and shower rate is found to be non-linear, suggesting additional 
corrections are needed for multiple-muon showers.
The measurement of $\alpha_\mathrm{T}$ 
was repeated after correcting the 
observed shower rates for the changes in 
pressure, but the difference in the new value was negligible.

The second method used to quantify the significance of the 
seasonal effect is the amplitude of a cosine fit to the data.
The temperature of the atmosphere is not 
expected to strictly track a cosine and, thus, neither does
the observed rate of events.
However, the modulations are expected to be periodic, 
with a period close to, but not exactly, one year, with higher 
temperatures in the summer and colder in the winter.
The amplitude of the fit demonstrates the 
strength of the effect independently from the used model.
Fig.~\ref{fig:moy} shows the average percent change in the multiple-muon rate as a function of month of the year, which is found to be more sinusoidal than the more finely binned version.
The function used for the fit in Fig.~\ref{fig:moy} is
\begin{linenomath*}
\begin{equation}
    f\left(t\right)=V_0+V\cos\left[\frac{2\pi}{T}\left(t-\phi\right)\right],
\end{equation}
\end{linenomath*}
where $V_0$ is the function average, $V$ is the relative change, 
$T$ is the period of modulation fixed at one year, and the 
phase $\phi$ is the time of maximum.
The best fit values are 
$V_0=0.00\pm0.01$\%,
$V=5.86\pm0.05$\%, and
$\phi=0.88\pm0.02$~months.
The phase implies a peak rate around January 27.
The value of $V$ gives a qualitative 
measurement for how much the multiple-muon 
showers rate varies throughout any given year.

The models used to measure the above 
quantities provide imperfect descriptions of the relationship between atmospheric conditions and the rate of 
multiple-muon air showers.
However, we can still consider how systematically 
changing the detector observables affects the measured 
quantities to determine if a systematic effect 
could give rise to the observed variations.
Here, we discuss the largest such effects.

The temperature and pressure measurements 
made by the ECMWF have an associated systematic 
uncertainty of $\pm$\SI{0.31}{K} and $\pm$\SI{1}{mbar}, 
respectively~\cite{minossingle}.
These uncertainties are the largest known systematic uncertainties in the measurement and are 
already included in the measurement of 
$\alpha_\mathrm{T}$ and $\beta$ in the fits above.

The detector data acquisition system writes data files in periods of up to \SI{2.5}{minutes}, 
depending on the data triggers operating at the time.
The detector events are written to the files in the order the trigger issues verdicts, so the data may appear out of order, and the files are closed when they reach a fixed file size~\cite{novasn}.
Occasionally, data at the end of one file and the start of the next will be misordered.
The livetime in such cases may be overestimated by as much as \SI{1}{s}.
Systematically increasing all livetimes by \SI{1}{s} has less than 0.5\% effect on the 
values of $V$ and $\alpha_\mathrm{T}$ and 1.5\% on the value of $\beta$.

The detector is a rectangular prism with length about 
four times its width and height.
Showers directed at the two smaller faces of the 
detector will have a smaller visible cross section of 
detector and will, thus, have lower multiplicities.
To account for this geometric effect, such showers 
have both their multiplicities and rate systematically increased by a factor of four.
The values of $\alpha_\mathrm{T}$ and $V$ decrease by less than 0.2\% and the value of beta increases by only 1\%.
Since the effect is small, this is not used to correct angular effects in the data.

The detector electronics are sensitive to the 
temperature and humidity of the atmosphere.
As a result, the electronics are noisier when the 
operating temperature is warmer, and up to 5\% more noise hits are observed in the 
summer months.
However, noise hits have much lower ADC counts than those made by the signal muon tracks and were not found to have any impact 
on the reconstruction of multiple-muon events.

None of the considered effects are large 
enough to have artificially created the observed 
winter maximum in the multiple-muon rate.
Additionally in the following section, these 
effects cancel when considering the relative 
measurements between the bins of multiplicity 
and the bins of zenith angle.

\section{Multiple-Muon Observables}
\label{sec:obs}

The two methods in the previous section 
demonstrate a clear seasonal variation with a peak during the winter.
As in the Near Detector analysis~\cite{ndpaper}, we 
observe that the strength of this effect changes under 
different shower observables that act as a 
proxy for the primary cosmic ray energy.
Here we examine the changes in the strength of the seasonal 
effect with the multiplicity and 
zenith angle of the shower.
The multiple-muon showers are divided into 
five sets depending on their multiplicities 
and zenith angles, respectively.
These divisions, shown in Fig.~\ref{fig:showers}, 
contain nearly equal numbers of showers.

\begin{figure}[t]
    \centering
    \includegraphics[width=0.48\textwidth]{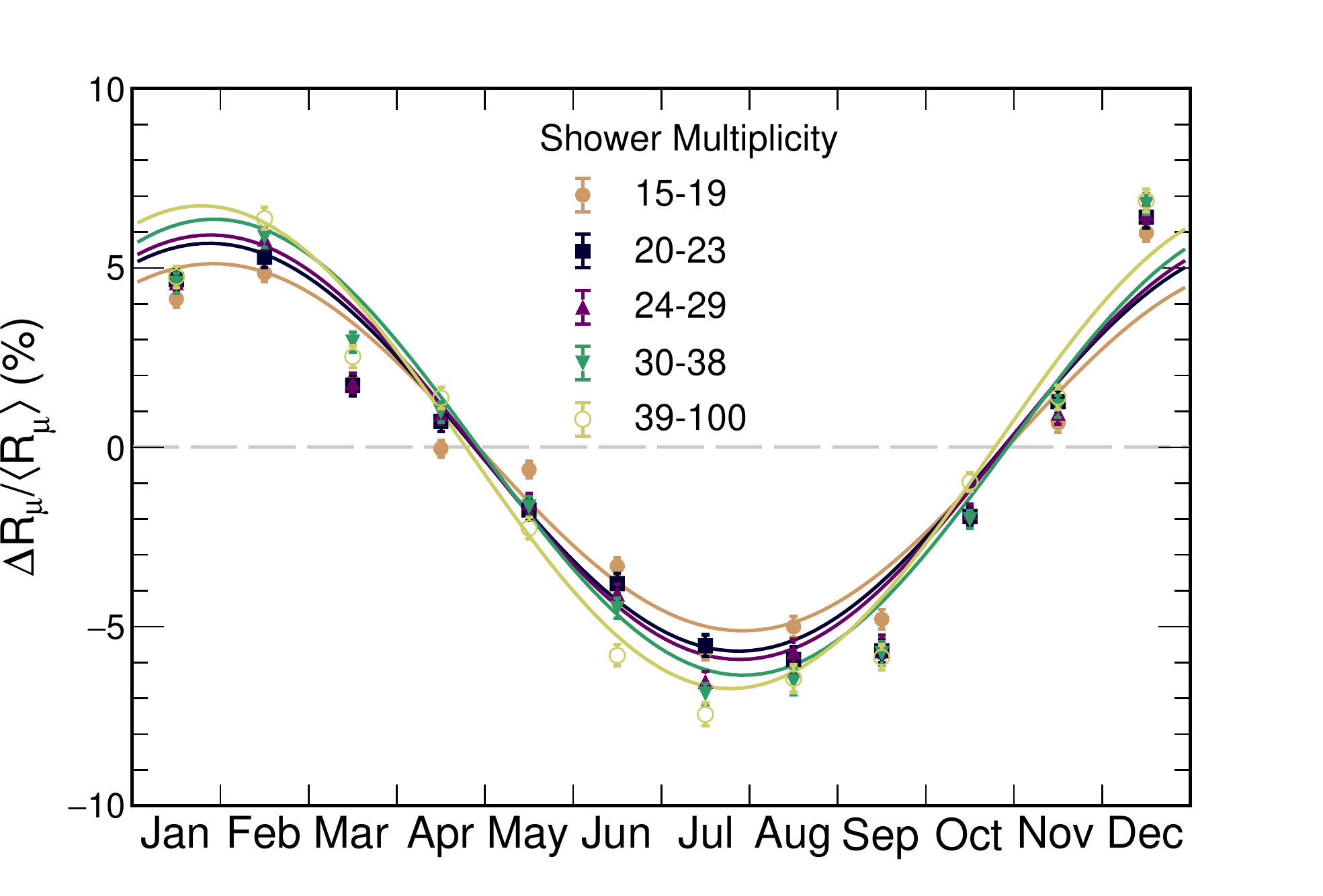}
    \caption{Percent variation in the multiple-muon data and a cosine fit for each of the five multiplicity bins.
    In each fit, the period is fixed at one year; all other parameters are unchanged.}
    \label{fig:cos}
\end{figure}

Fig.~\ref{fig:cos} shows the multiple-muon data and sinusoidal fit for each multiplicity bin as a function of month of year.
The sinusoidal fit amplitude, temperature coefficient, 
and barometric coefficient are measured 
within each multiplicity bin and reported in Table~\ref{tab:mult}.
Both the cosine fit amplitude and temperature coefficient demonstrate a stronger 
seasonal dependence at higher multiplicities.

\begin{table}[t]
\caption{The measured value of $V$, $\alpha_\mathrm{T}$, and $\beta$ for each of the 5 bins of measured muon multiplicity. The mean zenith angle for a given multiplicity is also shown.}
\begin{tabular}{c c c c c}
    \hline
    \hline
    Mult. & Mean $\cos\theta$ & V (\%) & $-\alpha_\mathrm{T}$ & $-\beta$ (\%/mbar) \\
    \hline
    15--19~ & 0.73 & $5.11 \pm 0.10$ & $0.94 \pm 0.04$ & $0.064 \pm 0.009$ \\
    20--23~ & 0.76 & $5.68 \pm 0.12$ & $1.07 \pm 0.05$ & $0.091 \pm 0.011$ \\
    24--29~ & 0.78 & $5.92 \pm 0.12$ & $1.08 \pm 0.05$ & $0.073 \pm 0.011$ \\
    30--38~ & 0.79 & $6.36 \pm 0.12$ & $1.10 \pm 0.05$ & $0.066 \pm 0.011$ \\
    39--100 & 0.82 & $6.73 \pm 0.13$ & $1.31 \pm 0.06$ & $0.054 \pm 0.012$ \\
    \hline
    15--100 & 0.78 & $5.86 \pm 0.05$ & $1.14 \pm 0.02$ & $0.077 \pm 0.005$ \\
    \hline
    \hline
\end{tabular}
\label{tab:mult}
\end{table}

We perform a similar analysis for each zenith angle bin.
The results are reported in Table~\ref{tab:zen}.
The bins of zenith angle nearest the 
horizon exhibit the strongest seasonal effect.
The table also shows the average 
multiplicity of showers within each bin.
The showers coming from nearest the 
horizon also have the lowest average multiplicities, 
which would be 
expected to exhibit the weakest seasonal 
variation from Table~\ref{tab:mult}.
We observe that the most vertical showers 
with the highest multiplicities exhibit 
the weakest seasonal change;
this is opposite what one would expect
based solely on the multiplicities.

\begin{table}[t]
\caption{The measured value of $V$, $\alpha_\mathrm{T}$, and $\beta$ for each of the 5 bins of zenith angle. The bins are sorted from vertical to horizontal.
The mean measured multiplicity of all showers in each bin is also shown.}
\begin{tabular}{c c c c c c}
    \hline
    \hline
    $\cos\theta$ & Mean Mult. & V (\%) & $-\alpha_\mathrm{T}$ & $-\beta$ (\%/mbar) \\
    \hline
    0.89--0.99 & 32.3 & $5.53 \pm 0.11$ & $0.95 \pm 0.04$ & $0.067 \pm 0.010$ \\
    0.83--0.89 & 29.0 & $5.70 \pm 0.12$ & $1.08 \pm 0.05$ & $0.095 \pm 0.011$ \\
    0.76--0.83 & 27.6 & $5.59 \pm 0.12$ & $1.01 \pm 0.05$ & $0.071 \pm 0.011$ \\
    0.66--0.76 & 26.4 & $5.96 \pm 0.12$ & $1.15 \pm 0.05$ & $0.065 \pm 0.011$ \\
    0.00--0.66 & 25.0 & $6.63 \pm 0.12$ & $1.24 \pm 0.05$ & $0.050 \pm 0.011$ \\
    \hline
    0.00--0.99 & 27.7 & $5.86 \pm 0.05$ & $1.14 \pm 0.02$ & $0.077 \pm 0.005$ \\
    \hline
    \hline
\end{tabular}
\label{tab:zen}
\end{table}

\section{Summary}

We observed that the rate of multiple-muon cosmic ray air showers in a detector near the surface presents a seasonal variation with a peak rate in the winter.
Additionally, we showed that this effect is dependent on the primary cosmic ray energy by looking at two detector observables, the multiplicity and zenith angle.
A stronger seasonal effect is seen for air showers with higher multiplicities or zenith angles near the horizon.
The amplitude of the seasonal modulation grows by 30\% from the lowest multiplicities to the highest multiplicities and by 20\% from the most vertical to the most horizontal showers considered.

For surface detectors where the threshold energy for detection is low,
the production altitude of muons in cosmic ray air showers is of the same 
magnitude as the muon decay length.
For example, a typical muon reaching the surface begins with \SI{5}{GeV} of energy at production and will traverse on average \SI{31.1}{km} before decaying.
Muon production occurs around \SIrange{15}{20}{km}, with higher altitudes in the summer months when the atmosphere is expanded.
Thus, the longer muon path length in summer months would give the produced muons a higher chance to decay before reaching the surface and reduce the number of observed muons.
The effects of particle decay on the seasonal rate of muons could be confirmed using Monte Carlo simulation such as CORSIKA~\cite{corsika}.
However, this effect is negligible in underground detectors where the muon energies are at least ten times larger and cannot explain all observations of seasonal variations for multiple-muon showers.

An interesting continuation of this study will be the inclusion of low multiplicity air showers from another detector trigger.
The two datasets could be combined to see if there is a threshold where the seasonal behavior for multiple-muons inverts as in underground detectors or flattens as expected for single-muons.
Additionally, comparisons to Monte Carlo simulation could be used to trace detector observables back to the primary cosmic ray energy.

This document was prepared by the NOvA collaboration using the resources of the Fermi National Accelerator Laboratory (Fermilab), a U.S. Department of Energy, Office of Science, HEP User Facility. Fermilab is
managed by Fermi Research Alliance, LLC (FRA), acting under Contract No. DE-AC02-07CH11359. This work
was supported by the U.S. Department of Energy; the
U.S. National Science Foundation; the Department of
Science and Technology, India; the European Research
Council; the MSMT CR, GA UK, Czech Republic; the
RAS, RFBR, RMES, RSF, and BASIS Foundation, Russia; CNPq and FAPEG, Brazil; STFC, UKRI, and the Royal Society, United Kingdom; and the state and University of
Minnesota. We are grateful for the contributions of the
staffs of the University of Minnesota at the Ash River
Laboratory and of Fermilab.

\bibliographystyle{apsrev4-2}
\bibliography{main}

\begin{thebibliography}{16}%
\makeatletter
\providecommand \@ifxundefined [1]{%
 \@ifx{#1\undefined}
}%
\providecommand \@ifnum [1]{%
 \ifnum #1\expandafter \@firstoftwo
 \else \expandafter \@secondoftwo
 \fi
}%
\providecommand \@ifx [1]{%
 \ifx #1\expandafter \@firstoftwo
 \else \expandafter \@secondoftwo
 \fi
}%
\providecommand \natexlab [1]{#1}%
\providecommand \enquote  [1]{``#1''}%
\providecommand \bibnamefont  [1]{#1}%
\providecommand \bibfnamefont [1]{#1}%
\providecommand \citenamefont [1]{#1}%
\providecommand \href@noop [0]{\@secondoftwo}%
\providecommand \href [0]{\begingroup \@sanitize@url \@href}%
\providecommand \@href[1]{\@@startlink{#1}\@@href}%
\providecommand \@@href[1]{\endgroup#1\@@endlink}%
\providecommand \@sanitize@url [0]{\catcode `\\12\catcode `\$12\catcode
  `\&12\catcode `\#12\catcode `\^12\catcode `\_12\catcode `\%12\relax}%
\providecommand \@@startlink[1]{}%
\providecommand \@@endlink[0]{}%
\providecommand \url  [0]{\begingroup\@sanitize@url \@url }%
\providecommand \@url [1]{\endgroup\@href {#1}{\urlprefix }}%
\providecommand \urlprefix  [0]{URL }%
\providecommand \Eprint [0]{\href }%
\providecommand \doibase [0]{https://doi.org/}%
\providecommand \selectlanguage [0]{\@gobble}%
\providecommand \bibinfo  [0]{\@secondoftwo}%
\providecommand \bibfield  [0]{\@secondoftwo}%
\providecommand \translation [1]{[#1]}%
\providecommand \BibitemOpen [0]{}%
\providecommand \bibitemStop [0]{}%
\providecommand \bibitemNoStop [0]{.\EOS\space}%
\providecommand \EOS [0]{\spacefactor3000\relax}%
\providecommand \BibitemShut  [1]{\csname bibitem#1\endcsname}%
\let\auto@bib@innerbib\@empty
\bibitem [{\citenamefont {An}\ \emph {et~al.}(2018)\citenamefont {An} \emph
  {et~al.}}]{dayabay}%
  \BibitemOpen
  \bibfield  {author} {\bibinfo {author} {\bibfnamefont {F.~P.}\ \bibnamefont
  {An}} \emph {et~al.} (\bibinfo {collaboration} {Daya Bay}),\ }\href
  {https://doi.org/10.1088/1475-7516/2018/01/001} {\bibfield  {journal}
  {\bibinfo  {journal} {JCAP}\ }\textbf {\bibinfo {volume} {1801}}\bibfield
  {number} {\bibinfo  {number} { (01)},\ \bibinfo {pages} {001}},\ }\Eprint
  {https://arxiv.org/abs/1708.01265} {arXiv:1708.01265 [physics.ins-det]}
  \BibitemShut {NoStop}%
\bibitem [{\citenamefont {De~Mendonça}\ \emph {et~al.}(2013)\citenamefont
  {De~Mendonça}, \citenamefont {Raulin}, \citenamefont {Echer}, \citenamefont
  {Makhmutov},\ and\ \citenamefont {Fernandez}}]{carpet}%
  \BibitemOpen
  \bibfield  {author} {\bibinfo {author} {\bibfnamefont {R.~R.~S.}\
  \bibnamefont {De~Mendonça}}, \bibinfo {author} {\bibfnamefont {J.~P.}\
  \bibnamefont {Raulin}}, \bibinfo {author} {\bibfnamefont {E.}~\bibnamefont
  {Echer}}, \bibinfo {author} {\bibfnamefont {V.~S.}\ \bibnamefont
  {Makhmutov}},\ and\ \bibinfo {author} {\bibfnamefont {G.}~\bibnamefont
  {Fernandez}},\ }\href {https://doi.org/10.1029/2012JA018026} {\bibfield
  {journal} {\bibinfo  {journal} {Journal of Geophysical Research: Space
  Physics}\ }\textbf {\bibinfo {volume} {118}},\ \bibinfo {pages} {1403}
  (\bibinfo {year} {2013})}\BibitemShut {NoStop}%
\bibitem [{\citenamefont {Adamson}\ \emph {et~al.}(2014)\citenamefont {Adamson}
  \emph {et~al.}}]{minossingle}%
  \BibitemOpen
  \bibfield  {author} {\bibinfo {author} {\bibfnamefont {P.}~\bibnamefont
  {Adamson}} \emph {et~al.},\ }\href
  {https://doi.org/10.1103/PhysRevD.90.012010} {\bibfield  {journal} {\bibinfo
  {journal} {Phys. Rev. D}\ }\textbf {\bibinfo {volume} {90}},\ \bibinfo
  {pages} {012010} (\bibinfo {year} {2014})},\ \Eprint
  {https://arxiv.org/abs/1406.7019} {arXiv:1406.7019 [hep-ex]} \BibitemShut
  {NoStop}%
\bibitem [{\citenamefont {Agafonova}\ \emph {et~al.}(2019)\citenamefont
  {Agafonova} \emph {et~al.}}]{opera}%
  \BibitemOpen
  \bibfield  {author} {\bibinfo {author} {\bibfnamefont {N.}~\bibnamefont
  {Agafonova}} \emph {et~al.} (\bibinfo {collaboration} {OPERA}),\ }\href
  {https://doi.org/10.1088/1475-7516/2019/10/003} {\bibfield  {journal}
  {\bibinfo  {journal} {JCAP}\ }\textbf {\bibinfo {volume} {1910}}\bibfield
  {number} {\bibinfo  {number} { (10)},\ \bibinfo {pages} {003}},\ }\Eprint
  {https://arxiv.org/abs/1810.10783} {arXiv:1810.10783 [hep-ex]} \BibitemShut
  {NoStop}%
\bibitem [{\citenamefont {Grashorn}\ \emph {et~al.}(2010)\citenamefont
  {Grashorn}, \citenamefont {de~Jong}, \citenamefont {Goodman}, \citenamefont
  {Habig}, \citenamefont {Marshak}, \citenamefont {Mufson}, \citenamefont
  {Osprey},\ and\ \citenamefont {Schreiner}}]{teff}%
  \BibitemOpen
  \bibfield  {author} {\bibinfo {author} {\bibfnamefont {E.}~\bibnamefont
  {Grashorn}}, \bibinfo {author} {\bibfnamefont {J.}~\bibnamefont {de~Jong}},
  \bibinfo {author} {\bibfnamefont {M.}~\bibnamefont {Goodman}}, \bibinfo
  {author} {\bibfnamefont {A.}~\bibnamefont {Habig}}, \bibinfo {author}
  {\bibfnamefont {M.}~\bibnamefont {Marshak}}, \bibinfo {author} {\bibfnamefont
  {S.}~\bibnamefont {Mufson}}, \bibinfo {author} {\bibfnamefont
  {S.}~\bibnamefont {Osprey}},\ and\ \bibinfo {author} {\bibfnamefont
  {P.}~\bibnamefont {Schreiner}},\ }\href
  {https://doi.org/10.1016/j.astropartphys.2009.12.006} {\bibfield  {journal}
  {\bibinfo  {journal} {Astropart. Phys.}\ }\textbf {\bibinfo {volume} {33}},\
  \bibinfo {pages} {140} (\bibinfo {year} {2010})},\ \Eprint
  {https://arxiv.org/abs/0909.5382} {arXiv:0909.5382 [hep-ex]} \BibitemShut
  {NoStop}%
\bibitem [{\citenamefont {Yurina}\ \emph {et~al.}(2019)\citenamefont {Yurina},
  \citenamefont {Bogdanov}, \citenamefont {Dmitrieva}, \citenamefont
  {Kokoulin},\ and\ \citenamefont {Shutenko}}]{decor}%
  \BibitemOpen
  \bibfield  {author} {\bibinfo {author} {\bibfnamefont {E.}~\bibnamefont
  {Yurina}}, \bibinfo {author} {\bibfnamefont {A.}~\bibnamefont {Bogdanov}},
  \bibinfo {author} {\bibfnamefont {A.}~\bibnamefont {Dmitrieva}}, \bibinfo
  {author} {\bibfnamefont {R.}~\bibnamefont {Kokoulin}},\ and\ \bibinfo
  {author} {\bibfnamefont {V.}~\bibnamefont {Shutenko}},\ }\href
  {https://doi.org/10.1088/1742-6596/1189/1/012010} {\bibfield  {journal}
  {\bibinfo  {journal} {J. Phys. Conf. Ser.}\ }\textbf {\bibinfo {volume}
  {1189}},\ \bibinfo {pages} {012010} (\bibinfo {year} {2019})}\BibitemShut
  {NoStop}%
\bibitem [{\citenamefont {Arunbabu}\ \emph {et~al.}(2017)\citenamefont
  {Arunbabu} \emph {et~al.}}]{grapes}%
  \BibitemOpen
  \bibfield  {author} {\bibinfo {author} {\bibfnamefont {K.}~\bibnamefont
  {Arunbabu}} \emph {et~al.},\ }\href
  {https://doi.org/10.1016/j.astropartphys.2017.07.002} {\bibfield  {journal}
  {\bibinfo  {journal} {Astropart. Phys.}\ }\textbf {\bibinfo {volume} {94}},\
  \bibinfo {pages} {22} (\bibinfo {year} {2017})}\BibitemShut {NoStop}%
\bibitem [{\citenamefont {Adamson}\ \emph {et~al.}(2015)\citenamefont {Adamson}
  \emph {et~al.}}]{minos}%
  \BibitemOpen
  \bibfield  {author} {\bibinfo {author} {\bibfnamefont {P.}~\bibnamefont
  {Adamson}} \emph {et~al.} (\bibinfo {collaboration} {MINOS}),\ }\href
  {https://doi.org/10.1103/PhysRevD.91.112006} {\bibfield  {journal} {\bibinfo
  {journal} {Phys. Rev.}\ }\textbf {\bibinfo {volume} {D91}},\ \bibinfo {pages}
  {112006} (\bibinfo {year} {2015})},\ \Eprint
  {https://arxiv.org/abs/1503.09104} {arXiv:1503.09104 [hep-ex]} \BibitemShut
  {NoStop}%
\bibitem [{\citenamefont {Acero}\ \emph
  {et~al.}(2019{\natexlab{a}})\citenamefont {Acero} \emph {et~al.}}]{ndpaper}%
  \BibitemOpen
  \bibfield  {author} {\bibinfo {author} {\bibfnamefont {M.~A.}\ \bibnamefont
  {Acero}} \emph {et~al.} (\bibinfo {collaboration} {NOvA}),\ }\href
  {https://doi.org/10.1103/PhysRevD.99.122004} {\bibfield  {journal} {\bibinfo
  {journal} {Phys. Rev.}\ }\textbf {\bibinfo {volume} {D99}},\ \bibinfo {pages}
  {122004} (\bibinfo {year} {2019}{\natexlab{a}})},\ \Eprint
  {https://arxiv.org/abs/1904.12975} {arXiv:1904.12975 [physics.ins-det]}
  \BibitemShut {NoStop}%
\bibitem [{\citenamefont {Tognini}(2018)}]{stefano}%
  \BibitemOpen
  \bibfield  {author} {\bibinfo {author} {\bibfnamefont {S.~C.}\ \bibnamefont
  {Tognini}},\ }\emph {\bibinfo {title} {{Observation of multiple-muon seasonal
  variations in the NOvA Near Detector}}},\ \href
  {https://doi.org/10.2172/1468447} {Ph.D. thesis},\ \bibinfo  {school}
  {Federal University of Goias} (\bibinfo {year} {2018})\BibitemShut {NoStop}%
\bibitem [{\citenamefont {Acero}\ \emph
  {et~al.}(2019{\natexlab{b}})\citenamefont {Acero} \emph {et~al.}}]{ana2019}%
  \BibitemOpen
  \bibfield  {author} {\bibinfo {author} {\bibfnamefont {M.}~\bibnamefont
  {Acero}} \emph {et~al.} (\bibinfo {collaboration} {NOvA}),\ }\href
  {https://doi.org/10.1103/PhysRevLett.123.151803} {\bibfield  {journal}
  {\bibinfo  {journal} {Phys. Rev. Lett.}\ }\textbf {\bibinfo {volume} {123}},\
  \bibinfo {pages} {151803} (\bibinfo {year} {2019}{\natexlab{b}})},\ \Eprint
  {https://arxiv.org/abs/1906.04907} {arXiv:1906.04907 [hep-ex]} \BibitemShut
  {NoStop}%
\bibitem [{\citenamefont {Ayres}\ \emph {et~al.}(2007)\citenamefont {Ayres}
  \emph {et~al.}}]{NOVATDR}%
  \BibitemOpen
  \bibfield  {author} {\bibinfo {author} {\bibfnamefont {D.~S.}\ \bibnamefont
  {Ayres}} \emph {et~al.} (\bibinfo {collaboration} {NOvA}),\ }\href
  {https://doi.org/10.2172/935497} {\bibinfo {title} {The {NOvA} technical
  design report}},\ \bibinfo {howpublished} {FERMI\-LAB-\allowbreak
  DESIGN-\allowbreak 2007-01} (\bibinfo {year} {2007})\BibitemShut {NoStop}%
\bibitem [{\citenamefont {Acero}\ \emph {et~al.}(2020)\citenamefont {Acero}
  \emph {et~al.}}]{novasn}%
  \BibitemOpen
  \bibfield  {author} {\bibinfo {author} {\bibfnamefont {M.~A.}\ \bibnamefont
  {Acero}} \emph {et~al.} (\bibinfo {collaboration} {NOvA}),\ }\href
  {https://doi.org/10.1088/1475-7516/2020/10/014} {\bibfield  {journal}
  {\bibinfo  {journal} {JCAP}\ }\textbf {\bibinfo {volume} {10}},\ \bibinfo
  {pages} {014}},\ \Eprint {https://arxiv.org/abs/2005.07155} {arXiv:2005.07155
  [physics.ins-det]} \BibitemShut {NoStop}%
\bibitem [{\citenamefont {Heck}\ \emph {et~al.}(1998)\citenamefont {Heck},
  \citenamefont {Knapp}, \citenamefont {Capdevielle}, \citenamefont {Schatz},\
  and\ \citenamefont {Thouw}}]{corsika}%
  \BibitemOpen
  \bibfield  {author} {\bibinfo {author} {\bibfnamefont {D.}~\bibnamefont
  {Heck}}, \bibinfo {author} {\bibfnamefont {J.}~\bibnamefont {Knapp}},
  \bibinfo {author} {\bibfnamefont {J.}~\bibnamefont {Capdevielle}}, \bibinfo
  {author} {\bibfnamefont {G.}~\bibnamefont {Schatz}},\ and\ \bibinfo {author}
  {\bibfnamefont {T.}~\bibnamefont {Thouw}},\ }\href@noop {} {\bibinfo {title}
  {{CORSIKA: A Monte Carlo code to simulate extensive air showers}}} (\bibinfo
  {year} {1998})\BibitemShut {NoStop}%
\bibitem [{\citenamefont {Dee}\ \emph {et~al.}(2011)\citenamefont {Dee} \emph
  {et~al.}}]{ecmwf}%
  \BibitemOpen
  \bibfield  {author} {\bibinfo {author} {\bibfnamefont {D.}~\bibnamefont
  {Dee}} \emph {et~al.},\ }\href {https://doi.org/10.1002/qj.828} {\bibfield
  {journal} {\bibinfo  {journal} {Quarterly Journal of the Royal Meteorological
  Society}\ }\textbf {\bibinfo {volume} {137}},\ \bibinfo {pages} {553–597}
  (\bibinfo {year} {2011})}\BibitemShut {NoStop}%
\bibitem [{\citenamefont {Bernero}\ \emph {et~al.}(2013)\citenamefont
  {Bernero}, \citenamefont {Olitsky},\ and\ \citenamefont
  {Schumacher}}]{muondecay}%
  \BibitemOpen
  \bibfield  {author} {\bibinfo {author} {\bibfnamefont {G.}~\bibnamefont
  {Bernero}}, \bibinfo {author} {\bibfnamefont {J.}~\bibnamefont {Olitsky}},\
  and\ \bibinfo {author} {\bibfnamefont {R.}~\bibnamefont {Schumacher}},\
  }\href {https://doi.org/10.1088/0954-3899/40/6/065203} {\bibfield  {journal}
  {\bibinfo  {journal} {J. Phys. G}\ }\textbf {\bibinfo {volume} {40}},\
  \bibinfo {pages} {065203} (\bibinfo {year} {2013})},\ \Eprint
  {https://arxiv.org/abs/1304.4945} {arXiv:1304.4945 [astro-ph.EP]}
  \BibitemShut {NoStop}%
\end{thebibliography}%

\end{document}